\documentclass[a4paper,12pt,oneside]{article}
\usepackage[cp1251]{inputenc}
\usepackage[english]{babel}
\usepackage{graphicx}
\usepackage{epsfig}
\usepackage{enumerate}
\usepackage{amsmath,amsfonts,amsthm}
\usepackage{amssymb}
\usepackage{fullpage}
\usepackage{cite}

\renewcommand{\frac}{\dfrac}

\title{Reverse Rotation of the Accretion Disk in RW Aur A:
Observations and a Physical Model}

\author{D.V. Bisikalo$^1 \footnote {E-mail: bisikalo@inasan.ru},$ A.V. Dodin$^2,$ P.V. Kaygorodov$^1,$ \\
S.A. Lamzin$^2,$ E.V. Malogolovets$^3,$ A.M. Fateeva$^1$}

\date{
$^1$ Institute of Astronomy, Russian Academy of Sciences,
Pyatnitskaya st. 48, Moscow, 109017 Russia, \\
$^2$ Sternberg Astronomical Institute, Moscow State University,
Universitetskii pr. 13, Moscow, 119991 Russia \\
$^3$ Special Astrophysical Observatory, Russian Academy of
Sciences, Nizhnii Arkhyz, Karachaevo–Cherkessia Republic, 369167
Russia
}

\begin{document}

\maketitle

\abstract{
Speckle interferometry of the young binary system RW Aur was performed with the 6-m
telescope of the Special Astrophysical Observatory of the Russian Academy of Sciences using filters with
central wavelengths of 550 nm and 800 nm and pass-band halfwidths of 20 nm and 100 nm, respectively.
The angular separation of the binary components was $1.448^{\prime\prime} \pm 0.005$ and the position angle of the system
was $255.9^o\pm 0.3.$ at the observation epoch (JD $2\,454\,255.9$). We find using published data that these values
have been changing with mean rates of  $+0.002{}^{\prime\prime}$/yr and  $+0.02{}^o$/yr, respectively, over the past 70 years. This
implies that the direction of the orbital motion of the binary system is opposite to the direction of the disk
rotation in RW~Aur~A. We propose a physical model to explain the formation of circumstellar accretion
disks rotating in the reverse direction relative to young binary stars surrounded by protoplanetary disks.
Our model can explain the characteristic features of the matter flow in RW~Aur~A: the high accretion rate,
small size of the disk around the massive component, and reverse direction of rotation.
}

\section*{Introduction}

T Tauri stars are young $(t<10^{7}$ years), low-mass
$(M\le 3\,M_\odot)$ stars in the stage of gravitational
contraction evolving toward the main sequence. Young
stars whose activity is due to the accretion of matter
from surrounding protoplanetary disks are classical
T Tauri stars (CTTS)~\cite{Bertout:1989}. Most CTTS are members
of binary and multiple systems~\cite{Mathieu:1994}, some of which
are surrounded by extended ring-like envelopes, i.e.,
outer (protoplanetary) disks. It is difficult to observe
the inner regions of protoplanetary disks bounding
the components due to the small angular sizes of
such systems. Although it has long been possible
to obtain structural images of the outer parts of the
protoplanetary disks in some T Tauri systems~\cite{Guilloteau-et-al:1999,Duvert-et-al:1998},
the structure of the inner parts of these disks has
only recently been imaged for a few wide binary systems~\cite{Hioki-et-al:2007, Mayama-et-al:2010}. The observations indicate that an area
of decreased density with a complex flow structure
including a circumstellar accretion disk forms
in the inner parts of the protoplanetary disks
surrounding young T Tauri stars.

The flow of matter in binary systems accreting material from
protoplanetary disks has been simulated numerically in many
studies~\cite{Artymowicz-Lubow:94,
Artymowicz-Lubow:96,Bate-Bonnell:97,Ochi-et-al:2005,Hanawa-et-al:2010,
Grinin-et-al:2010,Demidova-et-al:2010,Kaigorodov-et-al:2010,
Fateeva-et-al:2011,Sytov-et-al:2011,Val-Borro-et-al:2011}. The
simulation results indicate the following major flow elements in
these systems: circumstellar accretion disks around the system
components, a rarefied area in the central parts of the
protoplanetary disk adjoining the binary, and a complex system of
gas-dynamical flows. The analysis of calculation
results~\cite{Kaigorodov-et-al:2010, Fateeva-et-al:2011,
Sytov-et-al:2011} indicate that the supersonic orbital motion of
the system components through the gas of the protoplanetary disk
leads to bow shocks formation ahead of the circumstellar accretion
disk of each component; these bow shocks predominantly determine
the dynamics of the matter in the area of decreased density in the
central part of the protoplanetary disk, as well as accretion
processes in the system. The results of numerical simulations
agree well with the observed flow morphologies for systems whose
inner structure is resolved. However, at least one young binary,
RW~Aur, is known to have flow structures that seem to differ
principally from the results of currently available numerical
simulations.

RW~Aur is in the original list of T Tauri stars compiled by
Joy~\cite{Joy:1945} in 1945. One year earlier, the companion
RW~Aur B was discovered near the primary star RW~Aur~A at a
separation of $1.2-1.3^{\prime\prime}$~\cite{Joy-Biesbroeck:1944}.
Since the distance to RW~Aur is 140 pc~\cite{Bertout-et-al:2006},
the separation between the components exceeds 170~AU, even
projected onto the celestial sphere. According
to~\cite{Cabrit-et-al:2006}, the mass of RW~Aur~A is
$1.3-1.4M_\odot$ and the mass of RW~Aur B 0.7–0.9$M_\odot.$ Both
components are CTTS; according to~\cite{White-Ghez:2001}, the
accretion rate onto RW~Aur~A is $\sim 3\times10^{-7}$
M$_\odot$/yr, and the accretion rate onto RW~Aur~B a factor of ten
lower.

A bipolar jet is observed from
RW~Aur~A~\cite{Hirth-et-al:1994,Mundt-et-al:1998,Dougados-et-al:2000}.
Interaction between this jet and the surrounding gas can be traced
to distances of $\sim 4.6\times 10^4$ AU~\cite{McGroarty-Ray:2004}.
Arguments suggesting that the jet is inclined to the line of sight
by $45^o-60^o$ are given in~\cite{Cabrit-et-al:2006}.

A submillimeter image of the vicinity of RW~Aur with high angular
resolution was obtained in~\cite{Cabrit-et-al:2006}. Analysis of
the observational data indicates that RW~Aur~A is surrounded by an
axially symmetric gas–dust disk with an outer radius of $\sim 50$~AU.
The plane of the disk is perpendicular to the jet, and the
disk rotates clockwise viewed from the Earth. A spiral arm of
molecular gas curved counterclockwise goes out from the disk. The
length of this arm is $\sim 600$~AU. Molecular gas is observed
around RW~Aur~B as well, but its complex spatial and kinematic
structure make it unclear whether it forms a circumstellar disk.

It was established in~\cite{Cabrit-et-al:2006} that the direction of disk
rotation in RW~Aur~A  is opposite to the direction of
the axial rotation of the jet, which was first determined
in~\cite{Woitas-et-al:2005}. This raised doubts about widely accepted
mechanisms of jet formation in such systems. This
transformed an relatively ordinary description of one
particular, albeit interesting, binary system into a
fundamental problem that was very important to our
understanding of the physical processes in accretion
disks. It was recently concluded that the true
direction of jet rotation in RW Aur A cannot be determined
using available observations~\cite{Coffey-et-al:2012}. Nevertheless, the
system remains the subject of intense research be-
cause of the opposite directions of the rotation of the
circumstellar disk, which was determined with high
reliability, and the possible direction of orbital motion
proposed in~\cite{Cabrit-et-al:2006}. The direction of orbital rotation
was actually not measured in~\cite{Cabrit-et-al:2006}, but a
 particular direction was proposed to explain the formation
of the observed spiral arm due to the tidal effect of
RW Aur B, in accordance with theoretical
 considerations from~\cite{Clarke-Pringle:1993}.

Determining whether or not the system has a
unique reversed disk requires direct determination of
the direction of orbital rotation of the RW Aur binary
system. This paper presents observations indicating
that the angular rotation of the disk in RW Aur A
and the orbital angular momentum of the system are
indeed opposite, i.e. the disk rotation is reversed. The
results of three-dimensional, gas-dynamical
numerical simulations lead us to suggest a physical model
to explain the formation of reversed accretion disks
in young binary stars surrounded by protoplanetary
disks.

\section{Observations}

Speckle interferometry of RW Aur was performed
with the 6-m telescope of the Special Astrophysical
Observatory (SAO) of the Russian Academy of
Sciences on the night of December 15–16, 2011. A
system of speckle image detection based on a CCD with
internal signal amplification was used~\cite{Maksimov:2012}. Speckle
interferograms were obtained in the visible with filters
having central wavelengths of 550 and 800 nm and
pass-band halfwidths of 20 and 100 nm, respectively.
A plane achromatic micro-objective with 16-fold
 amplification provided a scale of 8.86$\pm$0.02 and
8.92$\pm$0.02 milliarcsec per pixel at the focal plane of the
telescope for the first and second filters, respectively.

The calibration of the image scales in the different
filters and the angular positioning were performed
using an opaque mask with two round holes placed
in the beam converging from the main mirror.
Observations of a bright single star using the mask
enabled the detection of interference bands from the
pair of apertures, which were used to derive the image
scale and position angle correction from the spatial
frequency and orientation of the bands. 2000 speckle
images were accumulated with exposure times of
20 ms in each filter, to determine the position
 parameters and magnitude differences. We estimated
the atmospheric seeing from the full width at half
maximum averaged over a run of speckle frames for a
stellar image; the seeing was $\approx 1^{\prime\prime}$ , slightly less than
the apparent component separation. Therefore, one-
quarter of the position angle was determined using a
long-exposure image of the binary.

The relative position of the components and the
difference in their magnitudes were determined using
power spectra averaged over a run of speckle
interferograms, without correction for the atmosphere-
transfer function~\cite{Labeyrie:1970}. The power spectrum for a
binary star is a cosine function whose period is
determined by the angular separation $\rho$ of the components,
and the band orientation by the position angle $\theta$ of the
two stars. The magnitude difference $\Delta m$ between the
components was calculated using the band contrast
in the power spectrum~\cite{Balega-et-al:2002,Pluzhnik:2005}.

Our analysis of speckle images obtained at epoch
2011.9580 (in Bessel years) in the 550 nm filter
yielded the angular distance between the components
$\rho = 1.448^{\prime\prime} \pm 0.005,$ the position angle of the binary
system $\theta=255.9^o\pm 0.3,$ and the magnitude
difference between the components $\Delta m=1.88^m \pm 0.04.$
The same values for the 800 nm filter are
$1.445^{\prime\prime} \pm 0.004,$ $255.9^o\pm 0.3,$ $1.43^m \pm 0.02$, respectively.

\section{Interpretation of the results}

The Table contains all previously measured values of $\rho$ and
$\theta$ for RW~Aur in various spectral ranges available from the
literature. The most accurate measurements were performed with the
Hubble Space Telescope using the Wide Field Camera and Planetary
Camera 2 in five filters (F336W, F439W, F555W, F675W, and F814W)
corresponding approximately to the U , B, V , R${}_c$ and
I${}_c$ filters~\cite{Ghez-et-al:1997}.

                     %%%%%%%%%%%%%%%%%%%%%%%%%%
\begin{table}
\begin{center}
\begin{tabular}{|c|c|c|c|c|}
\hline
Date &Filter & $\rho \pm \sigma_\rho,$ ${}^{\prime\prime}$
& $\theta \pm \sigma_\theta,$ ${}^o$ & Reference \\
\hline
1944.246 & V  & 1.17  & 253.8 & \cite{Joy-Biesbroeck:1944} \\
1944.248 & V  & 1.28  & 254.9 & \cite{Joy-Biesbroeck:1944} \\
1990.85  & K  & $1.39 \pm 0.03$ & $256 \pm 1.0$ & \cite{Ghez-et-al:1993} \\
1994.85  & U$^\prime$-I$^\prime_c$  & $1.4175 \pm 0.0034$ & $255.459 \pm 0.065$ & \cite{Ghez-et-al:1997} \\
1996.93  & K  & $1.397 \pm 0.026$ & $254.6 \pm 1.0$ & \cite{White-Ghez:2001} \\
1996.93  & L  & $1.396 \pm 0.018$ & $254.5 \pm 1.0$ & \cite{White-Ghez:2001} \\
1999.87  & N  & $1.42 \pm 0.02$ & $254.6 \pm 0.4$ & \cite{McCabe-et-al:2006}\\
1999.87  & IHW18 & $1.38 \pm 0.02$ & $254.7 \pm 0.4$ & \cite{McCabe-et-al:2006} \\
\hline
\end{tabular}
\end{center}
\caption{Values of $\rho$ and $\theta$ for RW~Aur~A$+$B available
from the literature.} \label{obs-data}
\end{table}
                     %%%%%%%%%%%%%%%%%%%%%%%%%%

\begin{figure}[ht]
 \begin{center}
\epsfig{file=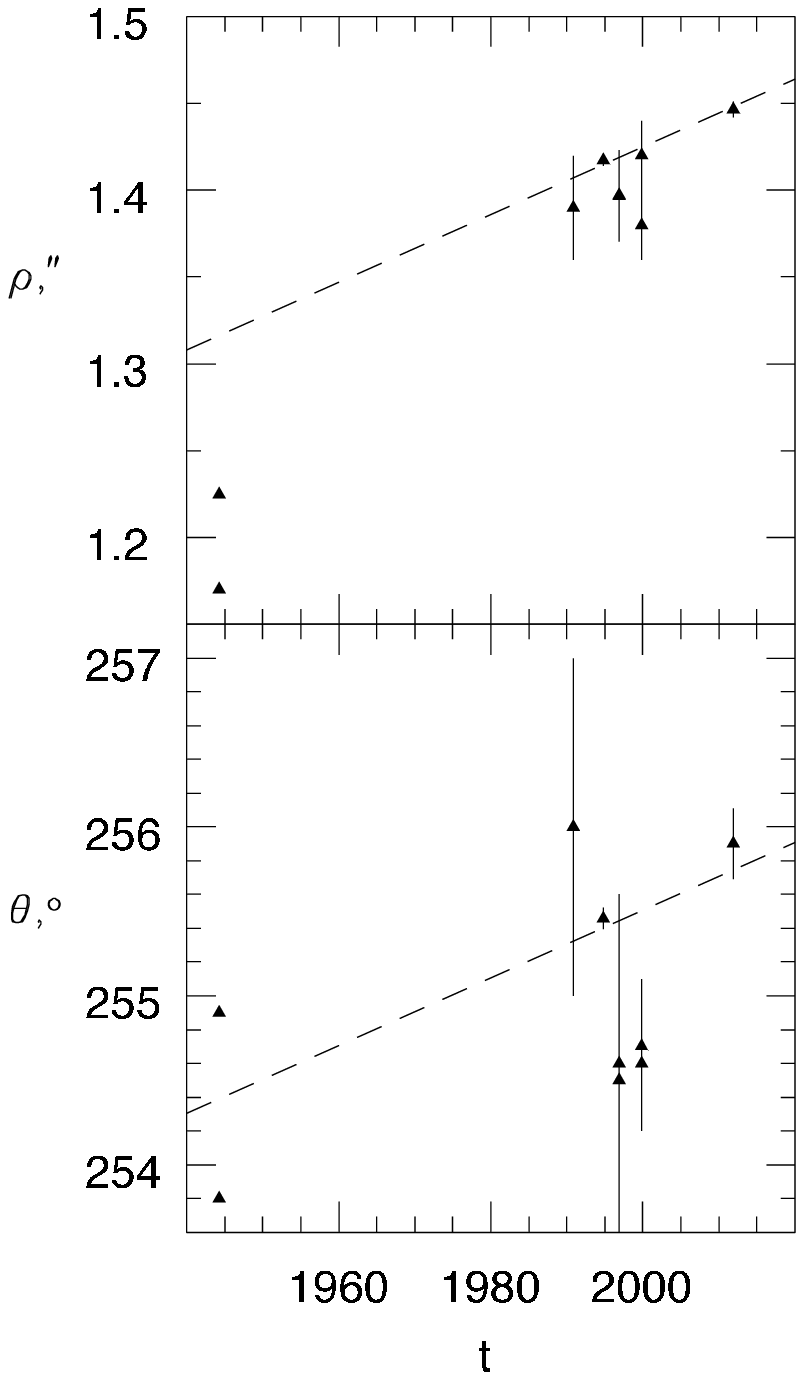,width=10cm}
  \caption{Angular distance between the components
(above) and position angle (below) of RW~Aur as functions of
time.}
  \label{fig-obs-data}
  \end{center}
\end{figure}

We plot the time variations of $\rho$ and $\theta$ in Figure~\ref{fig-obs-data},
using our data and those from the Table. The dashed
lines $y=kt+b$ are least-squares fits to the
observations. The parameters $(k_\rho,$ $b_\rho,$ and
$k_\theta,$ $b_\theta)$ of these lines
were calculated assigning weights that were inversely
proportional to the uncertainties $\sigma_\rho$ and $\sigma_\theta$.
The first two data points in~\cite{Joy-Biesbroeck:1944} do not have uncertainties;
therefore, we used their mean when calculating $k$ and $b$
and considered the rms deviation from the mean as
the uncertainty.

The coefficients $k_\rho$ and $k_\theta,$ characterize the mean
rates of variation of $\rho$ and $\theta$ over the $\approx 68$ years of
observation, and prove to be $+2.0\times 10^{-3}{}^{\prime\prime}$/yr and
$+2.0\times 10^{-2}{}^o$/yr, respectively.

For a distance to RW~Aur $\simeq 140$ pc, we find that the
tangential component of the velocity with which the components are
moving away from each other is $V_t \simeq 1.3$~km/s. If $\rho
\simeq 1.4^{\prime \prime},$ the component separation projected
onto the celestial sphere is  $r_t\simeq 200$ AU. Assuming that
the total mass of the components is $M_A+M_B<2.3\,M_\odot$~\cite{Cabrit-et-al:2006},
the condition that the
components be gravitationally bound requires that the relative
velocity of their motion be no greater than
$$
V_{max} = \sqrt{ \dfrac{2G \left(M_A + M_B \right)}{r} },
$$
where $r$ is the component separation. By substituting
the numerical value of rt (which is $\le r$) for $r$ in
this expression, we obtain $V_{max}\simeq 3.2$ km/s. This
suggests that the value of $V_t$ we have obtained
is consistent with the hypothesis that the
components RW Aur A and B form a gravitationally bound
system.

The available data are not sufficient to determine
the orbital parameters for this system. The orbital
period $P$ can be approximated using the inequality~$P \ge 2\rho/k_\rho \sim 1500$ yr.
Another estimate can be obtained from Kepler’s third law:
$$
P = \sqrt{ \dfrac{ a^3}{M_A+M_B } } > 700 \, \text{yr},
$$

assuming that the angular size of the semi-major axis $a$
of the orbit exceeds $1.4^{\prime \prime}/2=0.7^{\prime \prime}$.

Since $k_\theta > 0,$ the orbiting components move
counterclockwise when viewed from the Earth.
Therefore, the accretion disk in RW~Aur~A rotates
opposite to the direction of rotation of the binary
system. Accordingly, this suggests reverse rotation
of the accretion disk in RW~Aur~A.

            %%%%%%%%%%%%%%%%%%%%%
            %%%%%%%%%%%%%%%%%%%%%

\section{A model for the formation of a reversed accretion disk}

\begin{figure}[ht]
\center
\epsfig{file=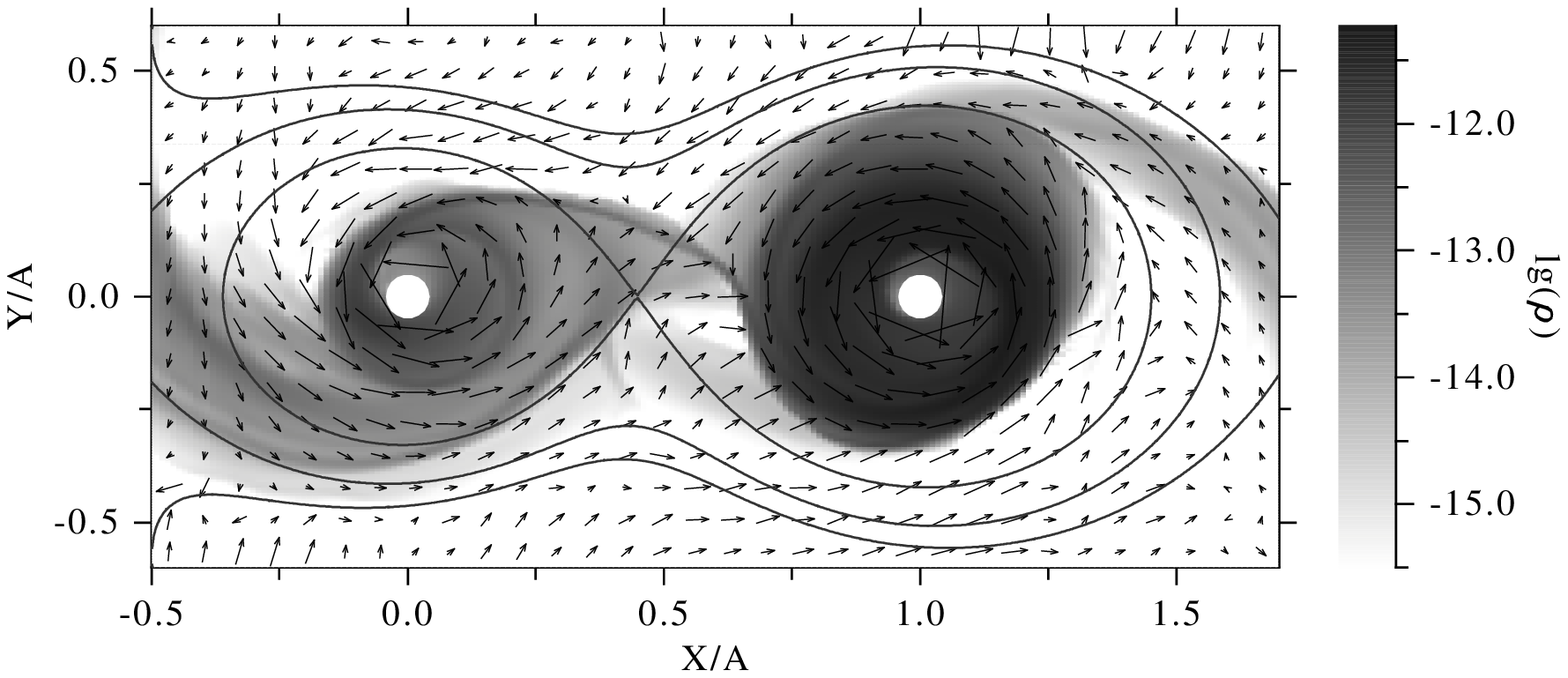,width=15.5cm}
\caption{Density and velocity distributions in the equatorial plane near the components of a young binary star. The binary rotates
counterclockwise. The more massive companion is to the right. The solid curves denote Roche equipotentials.}
\label{fig_old}
\end{figure}

\begin{figure}[ht]
\center \epsfig{file=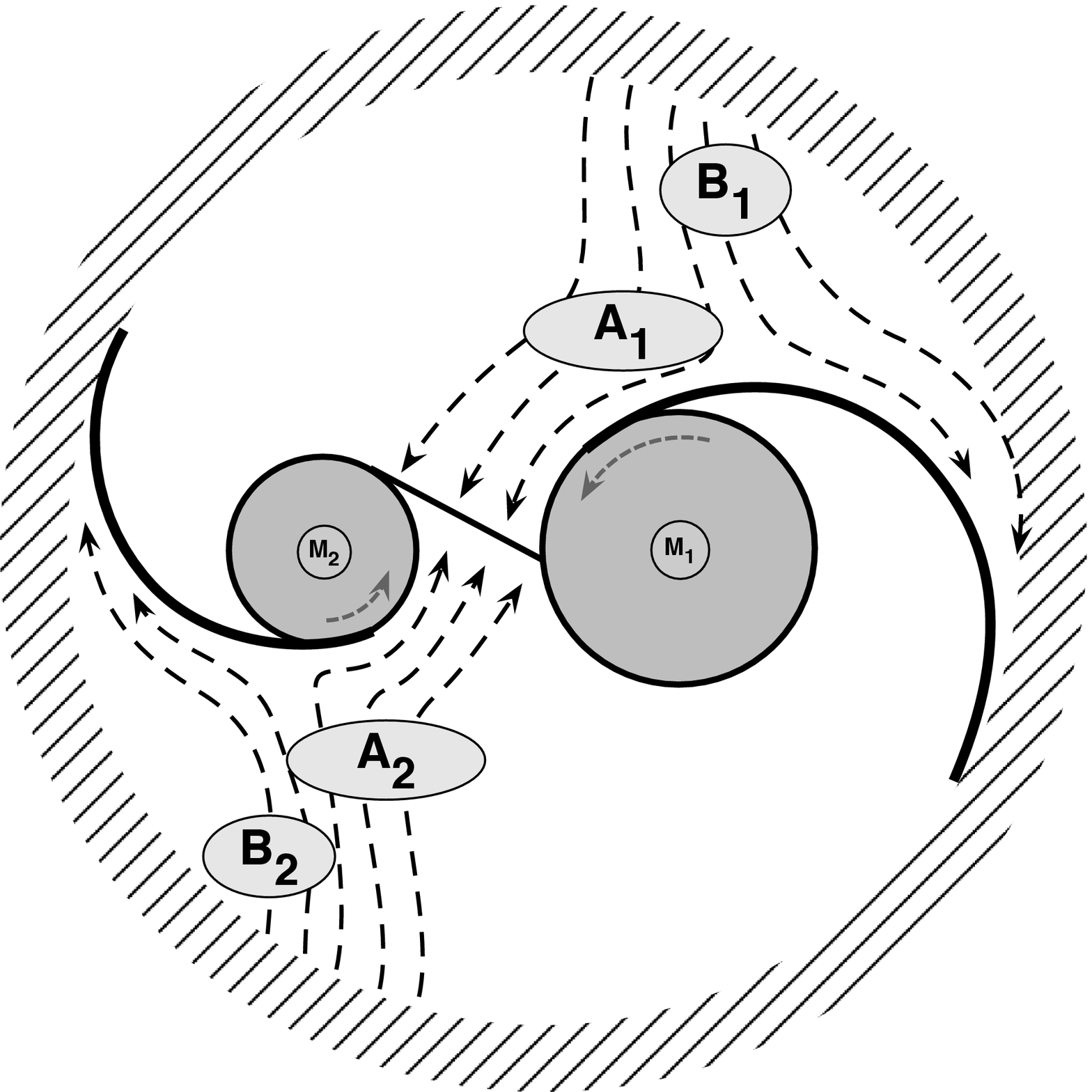,width=8cm} \caption{Schematic of the
flow structure in the inner parts of a protoplanetary disk around
a young binary star that rotates counterclockwise. The primary and
secondary are marked $\mathrm{M_1}$ and $\mathrm{M_2}$. The inner
edge of the protoplanetary disk is shown by the diagonal shading.
The accretion disks of the components are shown by dark grey
circles. Heavy lines denote the fronts of the outward moving
shocks that appear as the orbiting components move in the gas of
the circumbinary envelope. The dashed curves with arrows indicate
the directions of the main flows of matter in the system.}
\label{scm1}
\end{figure}

The direction of rotation of the accretion disk in RW~Aur~A is
defined by the angular momentum of the falling matter. For reverse
rotation to occur, the angular momentum delivered to the disk by
the falling matter must be opposite to the angular momentum of the
system. Numerical simulations of these systems have been performed
in various
studies~\cite{Artymowicz-Lubow:94,Artymowicz-Lubow:96,Bate-Bonnell:97,
Ochi-et-al:2005,Hanawa-et-al:2010,Grinin-et-al:2010,Demidova-et-al:2010,
Kaigorodov-et-al:2010,
Fateeva-et-al:2011,Sytov-et-al:2011,Val-Borro-et-al:2011}, have
not indicated the appearance of reversed accretion disks in binary
systems for a wide range of parameters. In addition, analysis of
the flow structure that forms in the vicinity of a young binary
indicates that flows directed opposite to the motion of the
components can exist~\cite{Fateeva-et-al:2011}. Indeed, the velocity field
presented in Figure~\ref{fig_old}
suggests that flows of matter in the
area between the components of a young binary move
opposite to the rotation of the circumstellar accretion
disks.

The main flow elements identified in~\cite{Fateeva-et-al:2011} are
shown schematically in Fig.~\ref{scm1}. The shaded area
corresponds to the inner edge of the protoplanetary disk; the
components of the system are marked $\mathrm{M_1}$ (the more
massive primary and $\mathrm{M_2}$ (the secondary); the dark grey
filled areas denote the circumstellar accretion disks, and the
solid curves show the bow shocks that form due to the supersonic
orbital motion of the stars through the envelope material. The
dashed curves with arrows denote the directions of the matter
motion. All the velocities are shown in a non-inertial coordinate
system that rotates together with the binary system
(counterclockwise).

Figure~\ref{scm1} shows that matter leaves the inner edge of the
protoplanetary disk and moves toward the binary components. Sooner
or later, it interacts with the bow shocks and divides into two
flow families, marked {\bf A} and {\bf B} (Fig.~\ref{scm1}). The
matter of flow family {\bf A} loses its angular momentum at the
shock and moves into the region between the components. There, the
flows collide and form a shock, which resembles a bar between the
circumstellar disks. The matter of flow family {\bf B} moves
toward the inner edge of the protoplanetary disk and carries away
excess of angular momentum. The bar is appreciably inclined, since
the orbital velocity of the low-mass component is greater (and the
bow shock is more powerful); accordingly, the flow of matter onto
this component is stronger.

\begin{figure}[ht]
\center
\epsfig{file=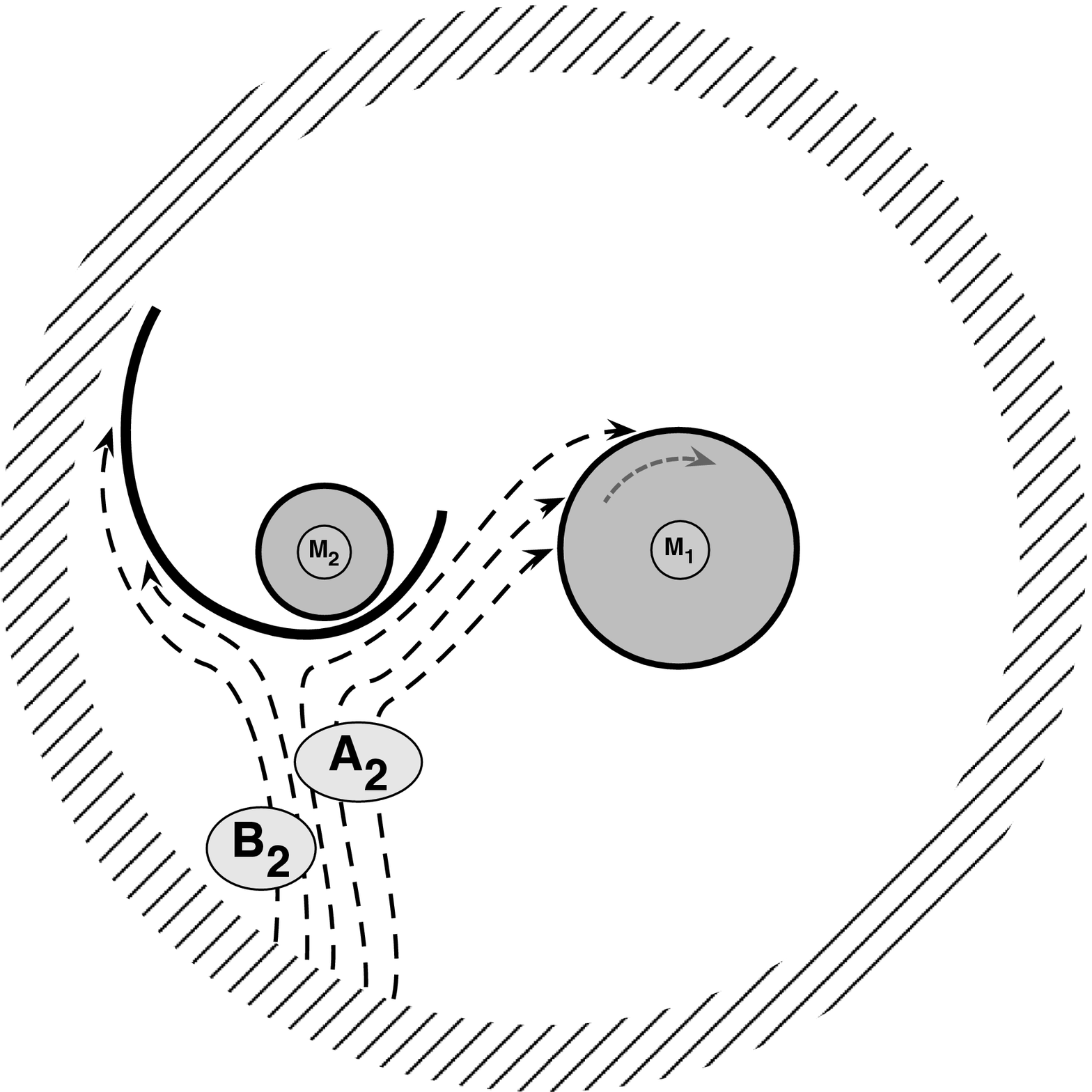,width=8cm}
\caption{Schematic of the flow structure in the inner parts
of the protoplanetary disk around a young binary star with
subsonic motion of the primary. The notation is the same
as in Fig.~\ref{scm1}.}
\label{scm2}
\end{figure}

The size of the semi-major axis of RW~Aur is not known exactly;
however, it is obviously greater than $\sim 200$~AU, since this is
the current observed component separation. We can estimate the
orbital velocities of the components to be $V_A\sim$~$1\text{km/s}$
and $V_B\sim 2\text{km/s}$, assuming that the orbit is circular
and the component masses are $0.8 M_\odot$ and $1.35 M_\odot$.
These estimates agree with the observed relative velocity of the
components, allowing for the projection of the motion onto the
celestial sphere. If the orbit is non-circular, the orbital
velocities of the stars will be functions of the phase, and (if
the system is not currently at apastron) they could be even
smaller. The bow shocks that determine the flow structure in these
systems can form only if the velocity of the components moving
through the gas of the common envelope exceeds the sound speed.
Assuming that the temperature of the surrounding gas is
$100\div200 K$~\cite{Cabrit-et-al:2006}, the sonic speed is
estimated to be $C_s\approx1.0\div1.6\text{km/s}$. It is obvious
that the velocity of the primary star in RW~Aur is close to (or
less than) the sonic speed. This suggests that the flow structure
in this system will differ dramatically from the structure
obtained in numerical simulations of the systems with supersonic
motion of the components.

In the solution in which the bow shock of the primary is either
absent or very weak, the matter flow due to the redistribution of
angular momentum at this shock ($\mathrm{\bf A_1}$ in
Fig.~\ref{scm1}) disappears, and the flow $\mathrm{\bf A_2}$ going
around the secondary will be completely captured by the more
massive component. The flow structure in this system is shown
schematically in Figure~\ref{scm2}. The flow $\mathrm{\bf A_2}$
(with angular momentum opposite to the orbital angular momentum)
collides with the accretion disk of the primary, which can lead to
the formation of a reversed accretion disk. Continuum observations
confirm the presence of a flow connecting the components of RW Aur
(see~\cite{Cabrit-et-al:2006}, Fig. 1a).

If the system is initially in a state in which one component moves
with a subsonic velocity (e.g., if the mass ratio is low), the
accretion disk of the primary will be reversed over the entire
lifetime. Accordingly, the star will always accrete matter with
negative angular momentum, and the direction of its intrinsic
rotation will also be reversed. However, if we assume that the
transition to a state with a reversed disk occurred not long ago,
the star should still retain its intrinsic angular momentum
(spin), whose direction coincides with that of the angular
momentum of the system and protoplanetary disk, since the star
accreted the matter with positive angular momentum before this
transition.

The accretion disk in such a system will have a number of
characteristic properties. The reversed rotation of the disk
relative to the stellar rotation implies that the angular momentum
in inner parts of the disk must partly annihilate, leading to a
substantial increase in the accretion rate. This effect is
especially appreciable when the star has a magnetic field (even if
it is weak). Moreover, the removal of angular momentum due to
annihilation should lead to an appreciable decrease in the disk
size, since it is not necessary for the disk to spread to the
radius of the last stable orbit defined by tidal interactions.

Analysis of available observational data confirms
both these hypotheses. Indeed, according to ~\cite{Cabrit-et-al:2006},
the radius of the accretion disk in RW~Aur~A is $\sim 50$ AU,
which is only $\sim 60\%$ of the radius of the last
stable orbit determined by Paczynski~\cite{Paczynski:77}.
The accretion rate onto this star appreciably exceeds
characteristic values for such systems, reaching $\sim 3\times10^{-7}$
M$_\odot$/yr~\cite{Cabrit-et-al:2006}.

Another important observational manifestation that determines the
validity of the model for reversed- disk formation could be
associated with the direction of rotation of the jet. As we noted
above, it is currently not possible to determine the direction of
jet rotation~\cite{Coffey-et-al:2012}. However, the formation of a
reversed disk can be explained in the proposed model without
violating generally accepted physical postulates, even if RW~Aur~A
has a jet that rotates opposite to the rotation of the accretion
disk~\cite{Woitas-et-al:2005, Cabrit-et-al:2006}.

Indeed, if the transition to the reversed-disk state
occured not long ago, the star should still have an
intrinsic angular momentum that coincides in
direction with the angular momentum of the system and
protoplanetary disk. The stellar magnetic field
(estimated to be $\simeq 1.5$ kGs for RW Aur A in~\cite{Dodin-et-al:2011}) should
lead to the capture of matter at the magnetosphere
boundary, accompanied by a corresponding change in
the direction of rotation of the matter. This results in
the formation of accretion columns, along with jets.
The jets can be accelerated by traditional mechanisms
in this case (see, for example,~\cite{Shu-et-al:1994,Najita-Shu:1994,Shu-et-al:1995,Ostriker-Shu:1995,Mohanty-Shu:2008,Shu-et-al2:1994}). The rotation
of the jets should coincide with the rotation of the star
(and, accordingly, the system), since the jets are fed
by matter captured by the magnetosphere. Therefore,
the direction of jet rotation should coincide with the
direction of system rotation in the proposed model,
although it may well be opposite to the rotation of the
accretion disk around a component.

\section*{Conclusions}

Our observations and the analysis of data obtained earlier have
enabled us to directly determine for the first time the true
direction of orbital rotation of the RW~Aur binary system. This
system displays an extremely interesting feature: reverse rotation
of the circumstellar disk around the massive component RW~Aur~A
relative to the orbital motion of the binary system.

We have proposed for the first time a mechanism to explain the
formation of reversed accretion disks in young binary systems
surrounded by protoplanetary disks. The secondary components in
some systems move in
 orbits with supersonic speeds relative to the gas of
the circumbinary envelope, while the more massive primary
components move with subsonic or
 approximately sonic speeds. The matter in such a system
passes through the bow shock that forms in front of the secondary
and divides into two flows, one of which can deliver matter with
``negative`` angular momentum onto the accretion disk of the
primary (Fig. 4). The disk that forms in this case will rotate
opposite to the orbital motion of the system and the intrinsic
rotation of the star.

A system can achieve a state when one component moves with
subsonic speed in several ways. First of all, the system may have
been in this state initially, but with a very low component-mass
ratio, as is more typical for star–planet systems than traditional
binary stars. The accretion disk of the primary will be reversed
in this system over the entire lifetime. The case when both
components originally moved with supersonic speeds, but one slowed
down at some time, is more interesting. This can occur as a result
of a change in the component-mass ratio as the components accrete
matter (since the accretion rate onto the more massive companion
is higher~\cite{Fateeva-et-al:2011}) or when the system approaches
apastron, if the orbital eccentricity is fairly large.

Numerical simulations show that both accretion disks have positive
angular momentum if the orbital motion is supersonic. As the
velocity of the primary decreases to the sonic speed, its bow
shock should disappear, after which its accretion disk will be fed
mostly by matter with negative angular momentum. As a result,
two zones can form for some time in the disk: an outer zone with
reversed rotation and an inner zone with direct rotation. The
angular momentum will annihilate at the boundary between these
zones, leading to an appreciable increase in the accretion rate.
The characteristic time required for the disk to be transformed
can be estimated using the ratio of the radius of the original
accretion disk (the radius of the last stable orbit) to the sonic
speed for the disk temperature. This time is $\sim 10^3$ years for
RW~Aur. If the system is transformed to a state with a reversed
disk as a result of a changing component-mass ratio, it cannot
leave this state in the future, since the component-mass ratio
will only increase with time. If such a transformation occurs when
the system approaches apastron, the star may demonstrate periodic
bursts in the accretion rate, which are, however, difficult to
observe, since all systems of this kind have long periods.

\bigskip
\bigskip

This work was supported by Basic Research Program P-21 of the
Presidium of the Russian Academy of Sciences “Non-Stationary
Phenomena in Objects of the Universe,” the Russian Foundation for
Basic Research (projects 11-02-00076, 11-02-01248, 12- 02-00047,
12-02-00393, and 12-07-00528), the Program of Support to Leading
Scientific Schools of the Russian Federation (NSh-3602.2012.2 and
NSh-5440.2012.2), a Grant of the President of the Russian
Federation in Support of Young Russian PhD Scientists
(MK-980.2012.2), and the Federal Targeted Program of the Ministry
of Education and Science of the Russian Federation “Research and
Scientific–Pedagogical Personnel of Innovational Russia” for years
2009–2013.

\end{document}